# Experimental study of the removal of excited state phosphorus atoms by $H_2O$ and $H_2$: implications for the formation of PO in stellar winds

Kevin M. Douglas[a], David Gobrecht[b], and John M.C. Plane[a*]

[a]*School of Chemistry, University of Leeds, Leeds, LS2 9JT, UK*

[b] *Department of Chemistry & Molecular Biology, University of Gothenburg, 40530 Göteborg, Sweden*

**Abstract.** The reactions of the low-lying metastable states of atomic phosphorus, P($^2$D) and P($^2$P), with $H_2O$ and $H_2$ were studied by the pulsed laser photolysis at 248 nm of $PCl_3$, combined with laser induced fluorescence detection of P($^2$D), P($^2$P) and PO. Rate coefficients between 291 and 740 K were measured, along with a yield for the production of PO from P($^2$D or $^2$P) + $H_2O$ of (35 ± 15)%. $H_2$ reacts with both excited P states relatively efficiently; physical (i.e. collisional) quenching, rather than chemical reaction to produced PH + H, is shown to be the more likely pathway. A comprehensive phosphorus chemistry network is then developed using a combination of electronic structure theory calculations and a Master Equation treatment of reactions taking place over complex potential energy surfaces. The resulting model shows that at the high temperatures within two stellar radii of a MIRA variable AGB star in oxygen-rich conditions, collisional excitation of ground-state P($^4$S) to P($^2$D), followed by reaction with $H_2O$, is a significant pathway for producing PO (in addition to the reaction between P($^4$S) and OH). The model also demonstrates that the PN fractional abundance in a steady (non-pulsating) outflow is under-predicted by about 2 orders of magnitude. However, under shocked conditions where sufficient thermal dissociation of $N_2$ occurs at temperatures above 4000 K, the resulting N atoms convert a substantial fraction of PO to PN.

## 1. Introduction

The recent observations discussed below indicate that the PO/PN ratio around oxygen-rich Asymptotic Giant Branch (AGB) stars ranges from ~1 to 20, and that the fractional abundances of PO and PN (i.e. with respect to the total gas which consists mostly of $H_2$, H and He) are of the order of $10^{-8}$ - $10^{-7}$. De Beck *et al.* (2013) postulated that PO and PN are the main carriers of phosphorus in the gas phase, with abundances up to several $10^{-7}$, since the solar elemental abundance of P is (2.6 ± 0.2) × $10^{-7}$ relative to H (Asplund *et al.* 2009). They also pointed out that chemical models could not account for these observations. For example, Gobrecht *et al.* (2016) modelled the chemistry around IK Tau, finding that PN was 6 × $10^{-7}$; however, the modelled PO was severely depleted with a fractional abundance of only 2 × $10^{-10}$.

Other recent observations of PO and PN around AGB stars include measured abundances of $1.7 \times 10^{-6}$ and $7.3 \times 10^{-7}$ respectively, relative to $H_2$ in IK Tau (Velilla Prieto *et al.* 2017). Ziurys *et al.* (2018) reported peak abundances in the oxygen-rich circumstellar envelopes for several AGB stars of $(0.5–1) \times 10^{-7}$ and $(1–2) \times 10^{-8}$, respectively, and concluded that the PO abundance is a factor of 5–20 greater than that of PN. This study suggests that phosphorus-bearing molecules are common in O-rich envelopes, and that a significant amount of phosphorus (>20%) remains in the gas phase.

PO and PN have also been observed in star-forming regions. For example, Lefloch *et al.* (2016) showed that PN arises from the outflow cavity, where the strong shock tracer SiO is produced. Radiative transfer analysis indicated a PO/PN ratio of ~3. The respective abundances of $2.5 \times 10^{-9}$ and $0.9 \times 10^{-9}$ imply a strong depletion, by approximately 2 orders of magnitude, of phosphorus in the quiescent cloud gas. This study also used shock modelling to demonstrate that atomic N plays a major role in the chemistry of PO and PN, concluding that the maximum temperature in the shock has to be larger than 4000 K. The production of these P-bearing species in shocks is supported by Fontani *et al.* (2019), who reported that the SiO and PN abundances are correlated over several orders of magnitude, and uncorrelated with gas temperature (which rules out alternative scenarios based on thermal evaporation from iced grain mantles). Mininni *et al.* (2018) found in a sample of nine massive dense cores that PN is well correlated with SiO in six out of the nine targets. However, in the other three objects the PN lines do not exhibit high-velocity wings, indicating that PN can be formed in colder and more quiescent gas through alternative pathways. Most recently, Bernal *et al.* (2021) observed in the Orion-KL nebula a PO/PN ratio of ~3, which is close to that measured in other warm molecular clouds. Finally, Rivilla *et al.* (2018) reported observations of PN and PO towards seven molecular clouds located in the Galactic Center. PN was detected in five out of the seven clouds whose chemistry appears to be shock-dominated. The two clouds where PN was not detected are exposed to intense radiation (UV, X-rays and cosmic rays). PO was detected only towards one cloud, with a PO/PN abundance ratio of ~1.5.

In response to these observations, a number of modelling studies of P-bearing species have been published in the last three years. Jimenez-Serra et al. (2018) showed how the measured PO/PN ratio can be used to constrain the physical conditions and energetic processing of the P source. These workers proposed that the reaction P + OH → PO + H, which had not been considered previously, could be an efficient source of gas-phase PO. Chantzos *et al.* (2020) modelled the depletion level of P in diffuse and translucent clouds, followed by Sil *et al.* (2021) who used chemical models to explore the evolution of P-bearing species in various environments including diffuse clouds and hot cores. These workers noted a significant anti-correlation between the PO/PN ratio and atomic N, with a ratio <1 in diffuse clouds where N is relatively high, and >1 in hot core/corino regions.

Souza et al. (2021) pointed out that these models of phosphorus-bearing molecules such as PO rely on rate coefficients set to those of the analogous reactions of NO, which is not a satisfactory state of affairs given the importance of PO and PN in pre-biotic chemistry (Schwartz 2006, Walton *et al.* 2021). Souza et al. (2021) therefore employed accurate multi-reference configuration interaction calculations on the N + PO and O + PN reactions to explore their underlying mechanisms and calculate potential energy barriers. The results confirmed previous assumptions that depletion of PO by N atoms is fast, with a branching ratio largely favoring O + PN rather than P + NO. Lastly, Molpeceres and Kästner (2021) published a theoretical study of the production of $PH_3$ on cold dust grains, resulting from the sequential addition of H atoms to adsorbed P.

In the present paper, we first present the results of an experimental study of the following reactions over the temperature range of 291 – 740 K involving the two low-lying excited (metastable) states of atomic phosphorus, $P(^2D)$ and $P(^2P)$, with $H_2O$ and $H_2$ ($H_2O$ is, after $H_2$ and CO, the third most abundant molecule in oxygen-rich conditions, see e.g. Williams and Hartquist (2013)):

| | | $\Delta H^\circ_{(0\ K)}$ (kJ mol$^{-1}$) | |
|---|---|---|---|
| $P(^2P) + H_2O$ | $\rightarrow PO + H_2$ | -339 | (R1a) |
| | $\rightarrow P(^2D) + H_2O$ | -88 | (R1b) |
| | $\rightarrow P(^4S) + H_2O$ | -224 | (R1c) |
| $P(^2D) + H_2O$ | $\rightarrow PO + H_2$ | -251 | (R2a) |
| | $\rightarrow P(^4S) + H_2O$ | -136 | (R2b) |
| $P(^2P) + H_2$ | $\rightarrow PH + H$ | -81 | (R3a) |
| | $\rightarrow P(^2D) + H_2$ | -88 | (R3b) |
| | $\rightarrow P(^4S) + H_2$ | -224 | (R3c) |
| $P(^2D) + H_2$ | $\rightarrow PH + H$ | +7 | (R4a) |
| | $\rightarrow P(^4S) + H_2$ | -136 | (R4b) |

where the reaction enthalpies (at 0 K) are calculated theoretically at the CBS-QB3 level of electronic structure theory (Montgomery *et al.* 2000). The role of $P(^2D)$ and $P(^2P)$, which are respectively 1.41 and 2.32 eV above the $P(^4S)$ ground state (Kramida *et al.* 2021), does not appear to have been considered previously in astrochemical models. However, our recent work on the P + $O_2$ reaction (Douglas *et al.* 2019), and the laboratory kinetics work described below (Section 3), shows that these metastable states are much more reactive than ground-state $P(^4S)$. In order to develop a comprehensive chemical network for phosphorus, we then estimate the rate coefficients for other important reactions of P-containing species by combining electronic structure calculations with Rice-Ramsperger-Kassel-Markus (RRKM) statistical rate theory, using a Master Equation formalism. Finally, the revised and extended phosphorus chemical network is incorporated into a stellar outflow model with pulsations (shocks), to explore whether the new network can model satisfactorily the absolute PO and PN abundances, and the PO/PN ratio, around a MIRA variable AGB star.

## 2. Experimental methods and numerical modelling

## *2.1 Reaction kinetics*

The experimental apparatus employed in this study has been discussed in detail elsewhere (Douglas et al. 2019, Douglas *et al.* 2020, Gómez Martín *et al.* 2009, Mangan *et al.* 2019), so only a brief overview is given here. All experiments were conducted in a slow-flow reactor (residence time of gas in the reactor ~ 1 s) using the pulsed laser photolysis-laser induced fluorescence (PLP-LIF) technique, with detection of either the first or second excited state of P (the $^2$D and $^2$P states respectively, hereafter collectively referred to as P$^*$), or PO. The reactor consists of a stainless-steel cell, with four horizontal side arms, orthogonally positioned, and a fifth vertical side arm. The cell is enclosed in a thermally insulated container, and can be operated at temperatures up to ~ 1000 K. Temperatures inside the reactor are monitored by two K-type thermocouples, located towards the centre of the reactor volume. The phosphorus radical precursor ($PCl_3$), reagent gases, and bath gas were introduced into the chamber *via* four of the five side arms, after being combined in a mixing manifold to ensure homogenous mixing. $PCl_3$ was introduced as a dilute mixture (between 1 and 5 % in either $N_2$ or He), with concentrations of $PCl_3$ in the reaction chamber typically being ≤ 0.2 % of the total flow. Flow rates were controlled using calibrated mass flow controllers (MKS instruments), with total mass flow rates ranging from 100 – 400 standard cm$^3$ min$^{-1}$. These total flow rates are sufficient to ensure a fresh flow of gas through the interaction region for each photolysis laser pulse. The pressure inside the reactor, as measured by a calibrated capacitance manometer (Baratron MKS PR 4000), ranged from ~ 3 to 11 Torr, and was controlled by a needle valve on the exit line to the pump. The photolysis and probe laser beams were introduced collinearly on opposite sides of the cell, and the fluorescence signal collected using a photomultiplier tube (PMT) (Electron Tubes, model 9816QB) mounted orthogonally to the laser beams. To increase the solid angle of collected fluorescence, a glass tube of ~1.5 cm diameter was positioned ~1 cm above the interaction region to act as a waveguide for transporting fluorescence photons along the vertical side arm to the PMT.

P$^*$ atoms were produced from the multiphoton photolysis of $PCl_3$ at 248 nm (Douglas et al. 2019) using a KrF excimer laser (Lambda Physik COMPEX 102):

$$PCl_3 + nh\nu \rightarrow P(^2P), P(^2D) + \text{co-products}$$

The excimer beam was loosely focused using a 50 cm focal length lens, with the focal point positioned approximately 10 cm beyond the interaction region, giving a beam cross section of ~ 8 mm$^2$ at the interaction region. Pulse energies at the interaction region ranged between 30 and 70 mJ pulse$^{-1}$. P$^*$ atoms were observed by time-resolved LIF spectroscopy, using the frequency doubled output of a Nd:YAG pumped dye laser (a Quantel Q-smart 850 pumping a Sirah Cobra-Stretch with a BBO doubling crystal). The transitions and wavelengths employed are listed in Table 1. Resonant fluorescence was measured by the PMT after passing through an appropriate interference filter (see Table 1), and recorded using a digital oscilloscope (LeCroy, LT262). The time delay between probe and photolysis laser beams was varied to produce scans of the relative P$^*$ concentration with time. A typical time-resolved LIF profile (Figure 2) typically consisted of 150 time steps and resulted from the average of between 5 and 10 individual delay scans.

**Table 1**. Transitions used for laser-induced fluorescence detection of the first two excited states of P and of PO.

| Radical Species | Excitation λ (nm)[a] | Transition | Laser dye | Filter[b] |
|---|---|---|---|---|
| $P(^2P)$[c] | 255.3 | $3s^23p^2(^3P)4s\ ^2P_{1/2} - 3s^23p^3\ ^2P°_{1/2}$ | Coumarin 503 | 254 (8) |
| $P(^2P)$[c] | 253.6 | $3s^23p^2(^3P)4s\ ^2P_{3/2} - 3s^23p^3\ ^2P°_{3/2}$ | Coumarin 503 | 254 (8) |
| $P(^2P)$[d] | 215.4 | $3s^23p^2(^1D)4s\ ^2D_{5/2} - 3s^23p^3\ ^2P°_{3/2}$ | Exalite 428 | 216 (10) |
| $P(^2P)$[d] | 215.3 | $3s^23p^2(^1D)4s\ ^2D_{3/2} - 3s^23p^3\ ^2P°_{1/2}$ | Exalite 428 | 216 (10) |
| $P(^2D)$ | 214.9 | $3s^23p^2(^3P)4s\ ^2P_{1/2} - 3s^23p^3\ ^2D°_{3/2}$ | Exalite 428 | 216 (10) |
| $P(^2D)$ | 213.6 | $3s^23p^2(^3P)4s\ ^2P_{3/2} - 3s^23p^3\ ^2D°_{5/2}$ | Exalite 428 | 216 (10) |
| PO | 246.3 | $A^2\Sigma^+ - X^2\Pi$ ($v',v''$ 0,0)[e] | Coumarin 503 | 254 (8) |

[a] BBO frequency-doubling crystal employed. [b] Interference filter peak transmission, FWHM in parentheses. [c] Transitions used to monitor loss of $P(^2P)$ with $H_2$ and $H_2O$. [d] Transitions used to compare LIF intensities with those of $P(^2D)$ lines. [e] $A^2\Sigma^+ - X^2\Pi$ ($v',v''$ 0,0) transition pumped at 246.3 nm and the non-resonant ($v',v''$ 0,1) LIF monitored at 255.4 nm (Sausa *et al.* 1986).

Two different methods were employed to introduce water vapour into the reaction cell. In the first, distilled $H_2O$ was freeze-pump-thawed to remove volatile contaminants, and the subsequent $H_2O$ vapour diluted in a glass bulb (in either $N_2$ or He) on a glass vacuum line. The dilute $H_2O$ vapour was then introduced to the chamber together with the other bath and precursor flows. For the second method, $H_2O$ vapour was entrained within a flow of either $N_2$ or He passing through a glass bubbler (a modified Dreschel bottle) containing distilled $H_2O$. The bubbler, which was maintained at a constant temperature using a water bath, was located before the relevant mass flow controller, and the gas flow exiting the bubbler shown to be saturated using a humidity probe (Process Sensing Technologies, PCMini52). The temperature of the bubbler was close to, or slightly below, room temperature. By measuring the temperature of the bubbler and the pressure of the $N_2$ / He gas over the $H_2O$, the concentration of the $H_2O$ entrained in the gas flow was determined *via* its known vapour pressure (Bridgeman & Aldrich 1964). The rate coefficients measure with these two methods for introducing $H_2O$ vapour into the reaction cell were in good agreement (within 7% at 300 K and 3% at 390 K); in practice, the second method was mostly used. When studying R1 and R2, the $H_2O$ gas flow was used to condition the reactor for 30 minutes prior to each experiment, to ensure that significant $H_2O$ vapour was not lost to the reactor walls during the kinetic measurements.

## 2.2 *PO Product Yields*

Experiments to determine the PO product yields from reactions R1 and R2 were carried out using the same slow-flow reactor and PLP-LIF technique describe above. PO was produced from the reaction of P* with $H_2O$ (R1a and R2a), and from the reaction of P* with any residual $O_2$ (R6a and R7a) in the reaction cell. P* atoms were produced *via* multiphoton photolysis of $PCl_3$ as described above, and PO monitored by time-resolved LIF spectroscopy using the $PO(A^2\Sigma^+ - X^2\Pi)$ transition (see Table 1). PO yields were determined by monitoring the amount of PO produced from reactions R1a and R2a as a function of $[H_2O]$.

| | | | |
|---|---|---|---|
| $P(^2P) + O_2$ | $\rightarrow PO + O$ | - 324 | (R6a) |
| | $\rightarrow P(^2D) + O_2$ | - 88 | (R6b) |
| | $\rightarrow P(^4S) + O_2$ | - 224 | (R6c) |
| $P(^2D) + O_2$ | $\rightarrow PO + O$ | - 236 | (R7a) |
| | $\rightarrow P(^4S) + O_2$ | - 136 | (R7b) |

To estimate the relative amounts of $P(^2D)$ and $P(^2P)$ produced following multiphoton photolysis of $PCl_3$ (for a particular precursor concentration and photolysis energy), the relative intensities of the LIF signal for the transitions in the 213.6 – 215.4 nm range were compared. As more $^2D$ than $^2P$ was produced in the multiphoton photolysis of $PCl_3$ (see Section 3.2), and as the $^2D$ state is efficiently relaxed by $N_2$ (Douglas et al. 2019), it was possible to tune the experimental conditions so that we were able to monitor PO produced primarily from either the $P(^2D)$ or the $P(^2P)$ state. For example, in experiments using $N_2$ as the bath gas, the high $[N_2]$ meant that the majority of $P(^2D)$ produced was removed by $N_2$ rather than by $H_2O$ (or $O_2$). Thus, as the $P(^2P)$ state is not efficiently relaxed by $N_2$ (Douglas et al. 2019), the majority of PO observed results from $P(^2P)$ reacting with $H_2O$ (or $O_2$), rather than $P(^2D)$. Conversely, in experiments using He as the bath gas, the majority of $P(^2D)$ produced was now removed by $H_2O$. As more $P(^2D)$ was produced in our experiments than $P(^2P)$, it is likely that the majority of PO observed was from the reaction of $P(^2D)$ with $H_2O$ rather than $P(^2P)$. As such, experiments measuring PO yields were carried out either using $N_2$ or He as the bath gas. For each bath gas, 3 or 4 sets of product yields were collected, over an $[H_2O]$ range of $0 - 1 \times 10^{16}$ molecule cm$^{-3}$ and $0 - 6 \times 10^{15}$ molecule cm$^{-3}$ for experiments in $N_2$ and He, respectively.

In experiments measuring the PO product yields, the recorded PO fluorescence signals were corrected for the quenching of the $PO(A^2\Sigma^+)$ by $H_2O$. This was done by determining the $PO(A^2\Sigma^+)$ fluorescence rate coefficient, $k_f$, as a function of $[H_2O]$. Using an oscilloscope, 100 fluorescence decays were averaged, and then fitted using an exponential function (Figure 1):

$$I_t = I_{f0}.\exp^{(-k_f.t)} \quad \text{(E1)}$$

By looking at the ratio of the fluorescence lifetime at a particular $[H_2O]$ compared to that with no $H_2O$ present, a correction factor, $S$, could be obtained, and applied to the absolute PO fluorescence signals obtained:

$$S = k_{f([H_2O])} \ / \ k_{f(H_2O=0)} \quad \text{(E2)}$$

In practice, $k_f$ was shown to be effectively independent of $[H_2O]$ in all experiments (inset to Figure 1), suggesting that quenching of the $PO(A^2\Sigma^+)$ signal was essentially all from the $N_2$ or He bath gas. The absolute PO LIF signals were therefore not correct for $H_2O$ quenching.

*Materials.* He (99.9995 %, British Oxygen Company (BOC)), $N_2$ (99.9995 %, BOC), and $H_2$ (99.99 %, BOC) were used without further purification. $PCl_3$ (≥ 99.0 %, VWR International

Ltd.) and deionized $H_2O$ were initially degassed by freeze-pump-thawing to remove volatile contaminants, and then made up as dilute vapours in $N_2$ or He.

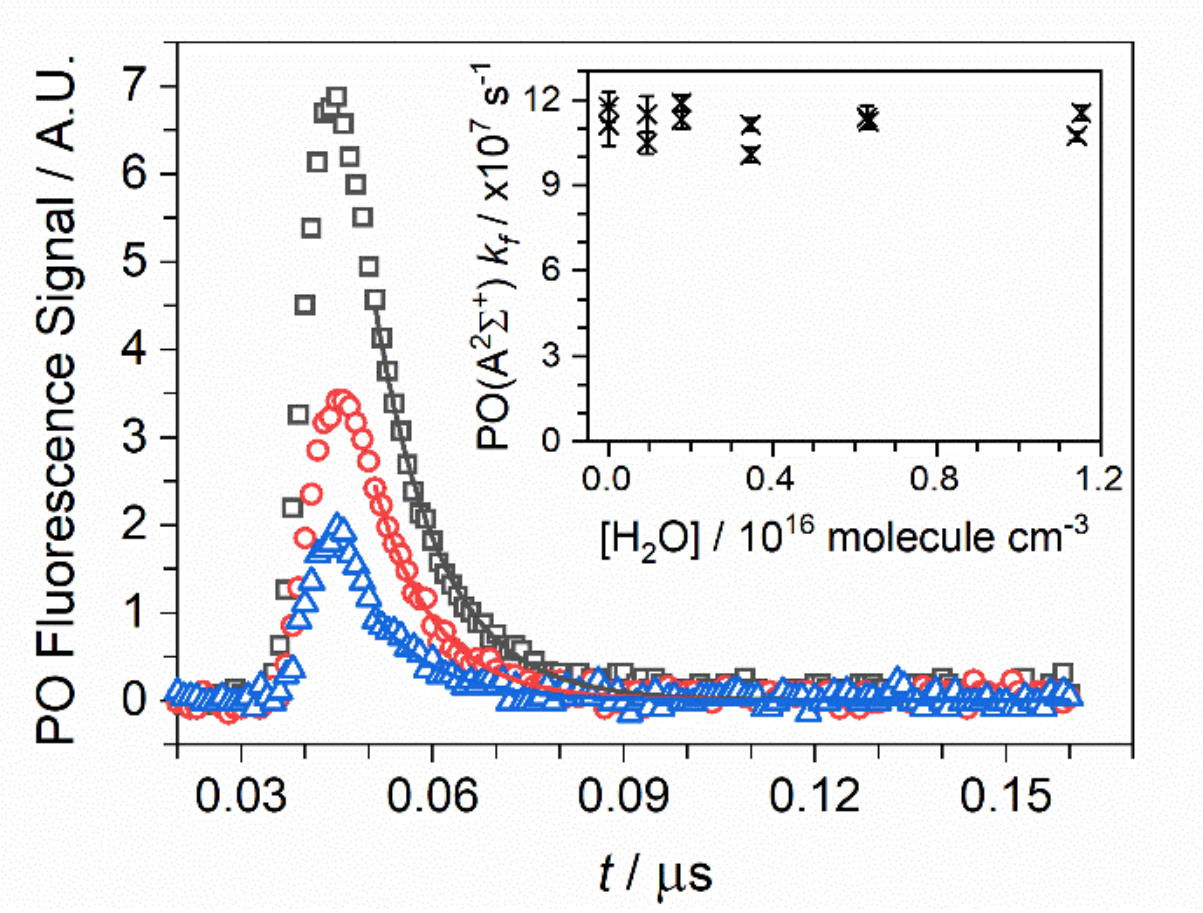

**Figure 1.** $PO(A^2\Sigma^+)$ fluorescence signal captured from the oscilloscope from experiments using $N_2$ as a bath gas, at three $[H_2O]$: black squares $[H_2O] = 1.8 \times 10^{16}$ molecule $cm^{-3}$; red circles $[H_2O] = 1.1 \times 10^{16}$ molecule $cm^{-3}$; blue triangles $[H_2O] = 0 \times 10^{16}$ molecule $cm^{-3}$. The solid lines are single exponential fits to each signal decay; these fits start far enough in time after the respective peaks in the bi-exponential signal that the early fast growth has ended. Inset: $PO(A^2\Sigma^+)$ fluorescence lifetime, $k_f$, *vs* $[H_2O]$, from experiments using $N_2$ as a bath gas.

## 3. Results

### *3.1 $P^*$ + $H_2O$ and $H_2$ Removal Rates*

An example of the time-resolved LIF signal for $P(^2D)$ in the presence of $H_2O$ can be seen in Figure 2. The LIF signal for both $P(^2D)$ and $P(^2P)$ decayed exponentially with time, with no increase in the LIF signal observed even at very short probe-photolysis delay times. As all experiments were carried out under pseudo first-order conditions (i.e. $[P^*] \ll [H_2O]$ or $[H_2]$), the loss of P* can be described by a single exponential of the form:

$$[P^*]_t = [P^*]_0 \, . \exp^{(-k'.t)} \quad \text{(E3)}$$

where $[P^*]_0$ is the initial concentration of P* from photolysis of $PCl_3$, $t$ is the time delay between the probe and photolysis laser pulses, and $k'$ is the experimentally observed pseudo first-order loss rate, which is equal to:

$$k' = k_r . [R] + k'_{\text{diff}} \quad \text{(E4)}$$

This expression encompasses the rates for all losses of P*, including diffusion and removal by the bath and precursor gases (summed as $k'_{\text{diff}}$), and removal by the co-reagent, R. Equation E3 was fitted to the $P^*$ profiles to extract the parameters $[P^*]_0$ and $k'$ (Figure 2). Plotting $k'$ *vs* [R] then yields a straight line with a gradient equal to the bimolecular rate constant, $k_r$, and the intercept equal to $k'_{\text{diff}}$ (inset Figure 2).

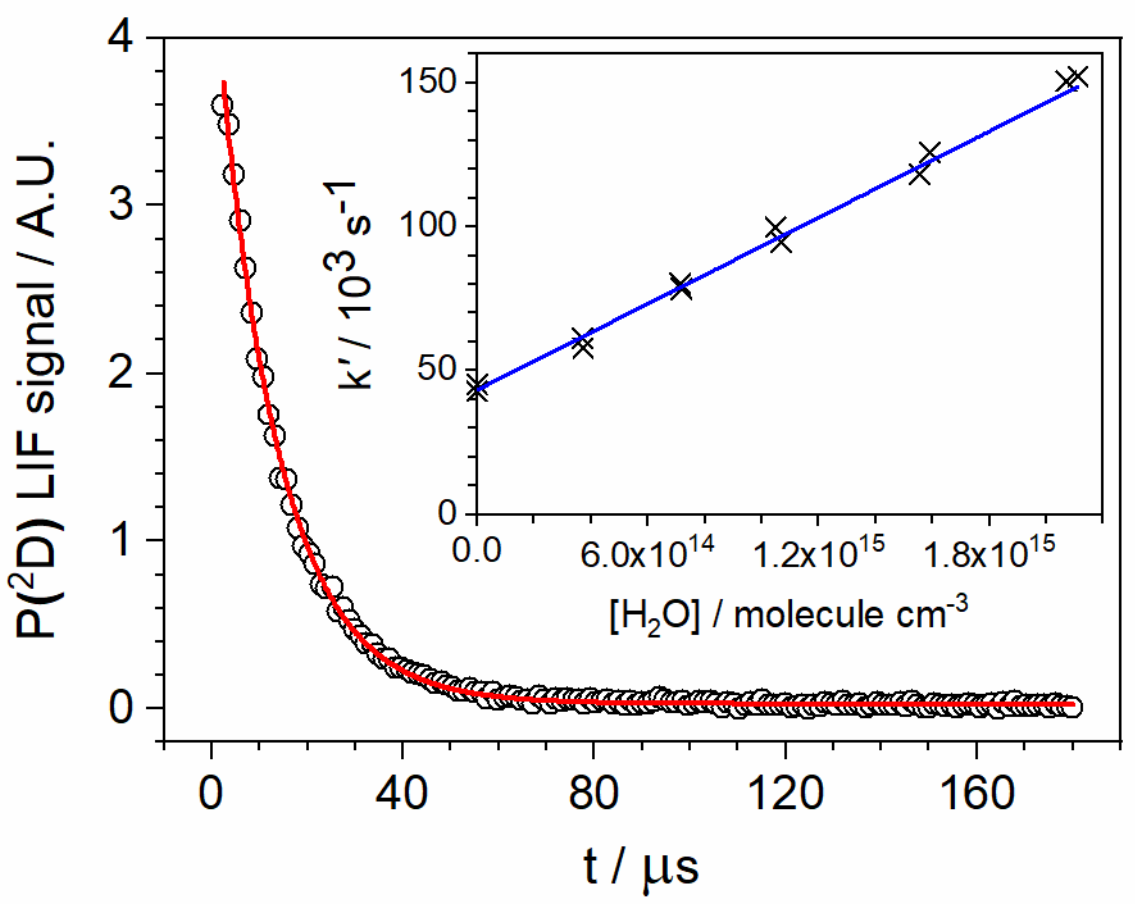

**Figure 2.** P($^2$D) LIF signal following PLP of $PCl_3$ at a total pressure of 7.64 Torr and [$H_2O$] = 7.19 × $10^{14}$ molecule cm$^{-3}$, at $T$ = 593 K. Inset: a bimolecular plot for reaction R2 at $T$ = 593 K, giving $k_2$ = (4.99 ± 0.30) × $10^{-11}$ cm$^3$ molecule$^{-1}$ s$^{-1}$.

The bimolecular rate coefficients for the removal of P$^*$ with $H_2O$ and $H_2$ are presented as a function of temperature in Figure 3, and listed in Table S1 in the Supporting Information (SI). Errors are reported at the 1σ level for the linear least-squares fitting of the pseudo-first order coefficients as a function of co-reagent concentration. The temperature range over which reactions R1 to R4 could be studied was limited by thermal decomposition in our reaction cell of the $PCl_3$ precursor at temperatures higher than ~750 K. No effects were observed on the bimolecular rate coefficients determined in this study when the $PCl_3$ precursor concentration and photolysis energy were varied by around a factor of 2, or when using different probe wavelengths. Varying the total pressure by around a factor of 3 also had no effect on the rate coefficients determined at room temperature.

There has only been one previous study investigating the removal of P$^*$ with $H_2$, with only a room temperature value reported (Acuña *et al.* 1973). In their study, P* was produced by single photon pulsed photolysis of $PCl_3$ at VUV wavelengths (λ < 160 nm), while removal of P* was observed using time-resolved atomic resonance absorption spectroscopy. As can be seen from Figure 3 and Table 1, the room temperature result for P($^2$P) + $H_2$ reported by Acuña et al. (1973) is around 70% larger than our value, with the two values lying just outside their mutual error limits. For P($^2$D) + $H_2$, our room temperature value is around 80% larger than that reported by Acuña et al. (1973). We reported a similar discrepancy between our room temperature values for the removal of P($^2$D) with $O_2$ and $CO_2$ (Douglas et al. 2019), with our values being around 50 and 80% faster, respectively, than those reported by Acuña et al. (1973). We are unable to account for these discrepancies; however, it should be noted that we obtain consistent room temperature values over a range of pressures, precursor concentrations, and photolysis energies, using two different probe wavelengths. Looking at the temperature dependence of P$^*$ + $H_2$, we see typical Arrhenius behaviour for both the P($^2$P) and P($^2$D) states, with a small positive temperature dependence, which can be parameterized as follows (see solid lines in Figures 3a and 3b; units: cm$^3$ molecule$^{-1}$ s$^{-1}$, 1σ errors):

$$k_{(\mathrm{P}(^2\mathrm{P}) + \mathrm{H_2})}(291 \leq T / \mathrm{K} \leq 740) = (6.50 \pm 0.18) \times 10^{-12} \exp^{[(-1056 \pm 14) / T]}$$

$$k_{(\mathrm{P}(^2\mathrm{D})+\mathrm{H_2})}(291 \leq T\,/\,\mathrm{K} \leq 684) = (8.04 \pm 0.48) \times 10^{-11} \exp^{[(-699 \pm 22)\,/\,T]}$$

There appear to have been no previous studies of the removal of $P^*$ with $H_2O$. As can be seen from Figure 3 and Table S1, the removal of both $P(^2P)$ and $P(^2D)$ by $H_2O$ shows only a weak temperature dependence between 300 and 700 K, and might effectively be temperature independent within error; this is in contrast to the removal of P* with other collision partners such as $O_2$, $H_2$, $CO_2$ and $N_2$, all of which show a moderate positive temperature dependence above 300 K (Douglas et al. 2019). Parameterization of the rate coefficients for $P^*$ + $H_2O$ over the experimental temperature range yields (see solid lines in Figures 3c and 3d; units: $cm^3$ $molecule^{-1}$ $s^{-1}$, 1σ errors):

$$k_{(\mathrm{P}(^2\mathrm{P})+\mathrm{H_2O})}(291 \leq T\,/\,\mathrm{K} \leq 599) = (1.06 \pm 0.06) \times 10^{-13} \exp^{[(-62 \pm 23)\,/\,T]}$$

$$k_{(\mathrm{P}(^2\mathrm{D})+\mathrm{H_2O})}(292 \leq T\,/\,\mathrm{K} \leq 706) = (5.01 \pm 0.22) \times 10^{-12} \exp^{[(-40 \pm 17)\,/\,T]}$$

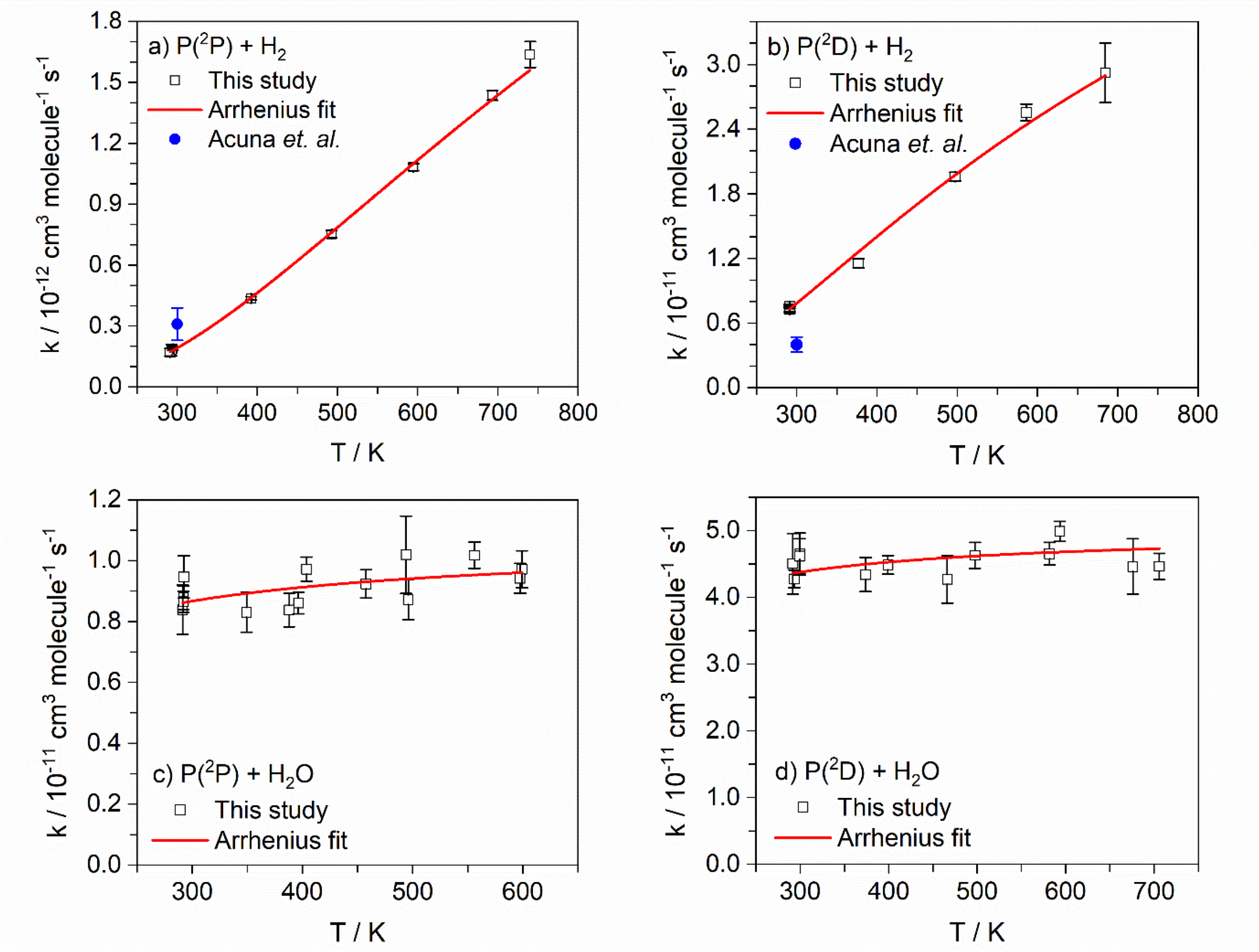

**Figure 3.** Temperature dependence of the rate coefficient for a) $P(^2P)$ + $H_2$, b) $P(^2D)$ + $H_2$, c) $P(^2P)$ + $H_2O$, and d) $P(^2D)$ + $H_2O$. Black open squares are from this study, with Arrhenius fits indicated by red lines; solid blue circles are from Acuña et al. (1973).

### *3.2 PO Yields*

PO LIF profiles produced following the multiphoton photolysis of $PCl_3$ in the presence of $H_2O$ are shown in Figure 4. Because the experiments were carried out under pseudo first-order conditions (i.e. $[P^*] << [H_2O]$ and $[O_2]$), the growth and loss of the PO signal can be described by a bi-exponential function of the form:

$$[\mathrm{PO}]_t = \gamma \left(\frac{k'_{\mathrm{growth}}}{k'_{\mathrm{loss}} - k'_{\mathrm{growth}}}\right) [\mathrm{P}^*]_0 \left(\exp^{-k'_{\mathrm{growth}} \cdot t} - \exp^{-k'_{\mathrm{loss}} \cdot t}\right) \quad \text{(E5)}$$

where $k'_{\mathrm{growth}}$ and $k'_{\mathrm{loss}}$ are the pseudo first-order rate coefficients for the reactions producing and removing PO, and $[P^*]_0$ is the amount of $P^*$ formed following multiphoton photolysis of the $PCl_3$ precursor, and $\gamma$ is the observed yield for PO production (as opposed to physical quenching). As PO itself does not react with $H_2O$, it is primarily lost through reaction with residual $O_2$ (R8):

$$PO + O_2 \rightarrow PO_2 + O \quad \text{(R8)}$$

Although the PO profiles are not strictly biexponential in nature, as the growth of the PO signal is from the reaction of two different species, they could be satisfactorily fit using equation E5 (see Figure 4), and the parameters $k'_{\mathrm{growth}}$, $k'_{\mathrm{loss}}$, and $\gamma$ extracted.

The PO yields obtained are plotted as a function of $[H_2O]$ in Figure 5. Because of day-to-day variations in the efficiency of the LIF collection system and power of the photolysis laser, the absolute PO yields collected under the same experimental conditions differed slightly. In order for them to be directly comparable to one another, and to be comparable to the PO yields obtained from the numerical simulations (see Section 3.3), they were put on a relative scale by dividing the yield at each $[H_2O]$ by the yield at $[H_2O] = 0$. In this manner, all PO yield *vs* $[H_2O]$ plots start at 1 and are directly comparable. As can be seen from Figure 5, the PO yields obtained using both $N_2$ and He as a bath gas increase with increasing $[H_2O]$, indicating P* does indeed react with $H_2O$ to produce PO.

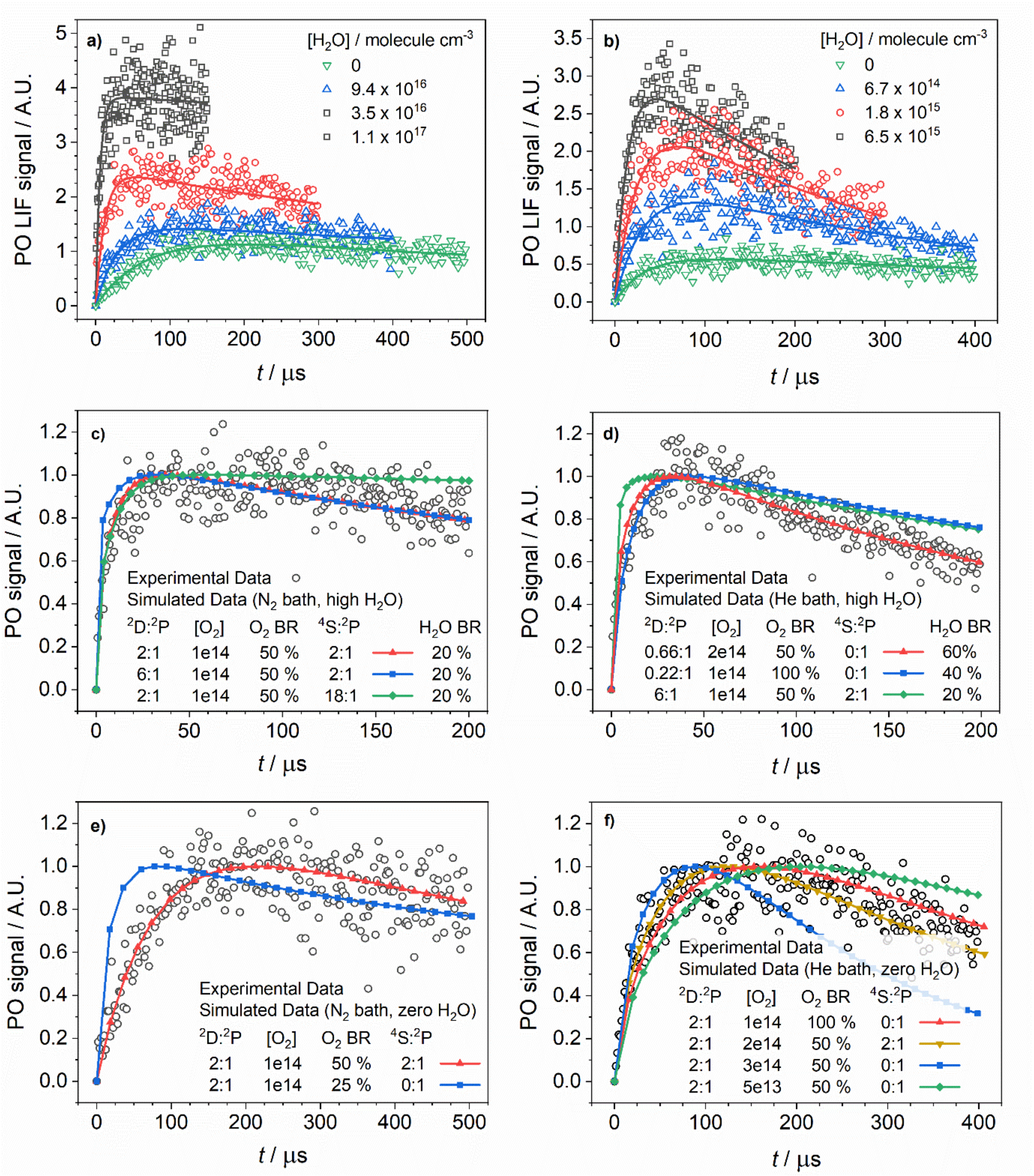

**Figure 4.** Top panels: PO LIF signals following PLP of $PCl_3$ at various [$H_2O$]. Solid lines are fits of Equation 5 to the data. Middle panels: Simulated and experimental PO profiles at high $H_2O$ ([$H_2O$] = $6 \times 10^{15}$ (left side) and [$H_2O$] = $1 \times 10^{15}$ (right side)). Bottom panels: Simulated and experimental PO profiles at zero $H_2O$. All left-hand panels are for experiments in He, while all right-hand panels are for experiments in $N_2$. Simulated fits which are considered good are (poor fits and reasons in brackets): c) red upward triangles (blue squares rise too fast, green diamonds loss too slow); d) red upward triangles (blue squares loss too slow, green diamonds rise too fast and loss too slow); e) red upward triangles (blue squares rise too fast); f) red upward triangles and yellow downward triangles (blue squares loss too fast, green diamonds rise and loss too slow).

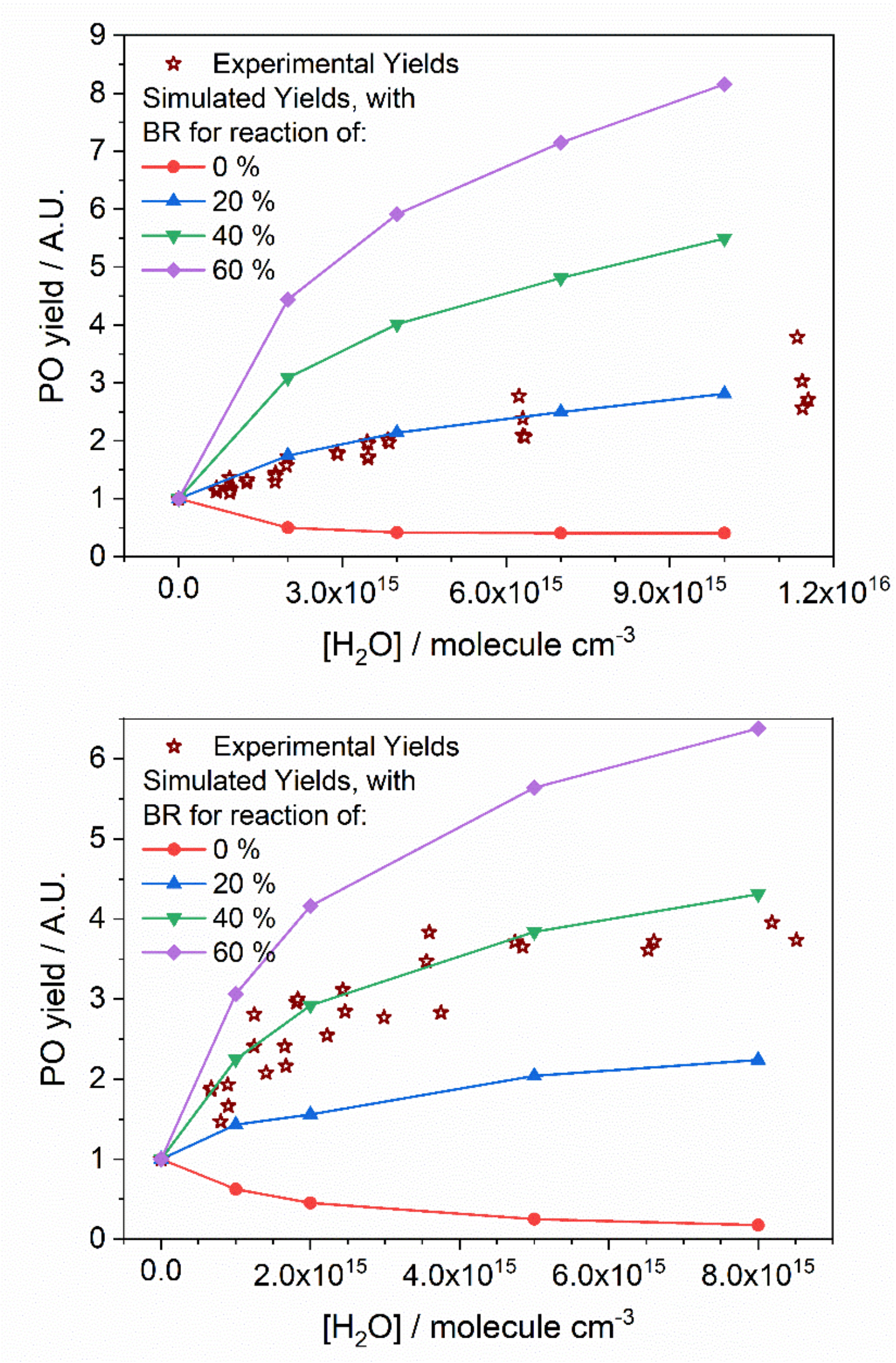

**Figure 5.** Experimental (open dark red stars) and simulated (closed symbols) PO yields. Top panel: experiments carried out in He, with model input parameters of $^2D/^2P = 2:1$, $[O_2] = 1 \times 10^{14}$ cm$^{-3}$, $O_2$ BR = 100%, $^4S:^2P = 0:1$. Bottom panel: experiments carried out in $N_2$, with model input parameters of $^2D/^2P = 2:1$, $[O_2] = 1 \times 10^{14}$ cm$^{-3}$, $O_2$ BR = 50%, $^4S:^2P = 2:1$.

### *3.3 Numerical Simulations and Determination of PO Branching Ratios*

The PO product yields ($\gamma$) do not themselves give the branching ratios (BRs) for PO formation from reactions R1 and R2. Instead, to determine the BR from the product yield, the other processes which remove P* need to be accounted for. Table S2 gives the rate coefficients of the main reactions removing P* in our experiments, together with the concentrations of the species employed in the experiments run with either $N_2$ or He as a bath gas. When no $H_2O$ was present in the reactor, P* could be removed by the $PCl_3$ precursor, any residual $O_2$, and in the case of P($^2$D), any $N_2$ present. Knowing the concentrations of these species, the total first-order removal of each state of P* can be estimated, as well as the percentage of P* that is removed by each species. Assuming reaction of P* with $O_2$ only produces PO (rather than relaxing the P*), we can assign the absolute PO yield observed in this experiment to the fraction of P* reacting with $O_2$ (this is 7.0% for the experiments using $N_2$ bath gas, see Table S2). In an experiment where $H_2O$ is present, P* can now be removed by the $PCl_3$ precursor, the residual $O_2$, and the $H_2O$ (and in the case of P($^2$D), any $N_2$ present). Again, knowing the concentrations of these species we can determine the total first-order removal rate of each state of P*, and determine the percentage of P* that is removed by each species. In experiments using $N_2$ bath gas, and with an [$H_2O$] of $1 \times 10^{16}$ cm$^{-3}$, 0.9% of the P* is removed by $O_2$, and 47.5% by $H_2O$. If the removal of P* with $H_2O$ proceeds only *via* the reactive channel producing PO, then 48.4% of the P* will now go on to produce PO, compared with only 7.0% when no $H_2O$ is present. As such the absolute PO yield observed will be 6.9 times higher (47.5 / 7.0) at high $H_2O$ compared to zero $H_2O$. Thus, by comparing the PO yield at various [$H_2O$] relative to that with [$H_2O$] = 0, and knowing the percentage of P* that is being removed by $H_2O$ and $O_2$, BRs for the reaction of P* with $H_2O$ can be determined.

In practice, extracting a BR directly from the PO yields is complicated by several factors, which are discussed in the SI. Numerical simulations were therefore carried out using the numerical integration package Kintecus (Ianni 2003). PO profiles, over a range of [$H_2O$] and for a range of P* + $H_2O$ BRs, were simulated and fit using Equation E5, and simulated PO yields obtained. These simulations were carried out over a range of input parameters as shown in Table 2. For a particular set of input parameters, the BR for PO production could be determined by comparing the simulated PO yields for a range of BRs to the experimentally determined yields. Figure 5 compares the PO yields obtained when using $N_2$ and He as a bath gas with two sets of simulated yields. As can be seen, the input parameters used in the top panel ($^2$D/$^2$P = 2:1, [$O_2$] = $1 \times 10^{14}$ cm$^{-3}$, $O_2$ BR = 100%, $^4$S:$^2$P = 0:1) predict a BR of ≈ 35%, while the input parameters in the bottom panel ($^2$D/$^2$P = 2:1, [$O_2$] = $1 \times 10^{14}$ cm$^{-3}$, $O_2$ BR = 50%, $^4$S:$^2$P = 2:1) predict a BR of ≈ 20%. Varying the model input parameters over the ranges given in Table 2 resulted in the predicted BR ranging from around 5% to over 100% (i.e. unphysical). However, many of the sets of input parameters could be ruled out by direct comparison of the simulated PO traces with the experimental traces. Figure 4 compares experimental PO traces produced at high and zero [$H_2O$] with simulated traces produced under the same conditions. If either the simulated high or zero $H_2O$ trace produced for a particular set of model conditions did not match with the experimental trace, that set of conditions was ruled out. For example, in Figure 4e the set of conditions that produced the PO trace shown by the blue squares clearly does not match the experimental PO trace, and as such, that particular set of conditions ($^2$D/$^2$P = 2:1, [$O_2$] = $1 \times 10^{14}$ cm$^{-3}$, $O_2$ BR = 25 %, $^4$S:$^2$P = 0:1) can be ruled out. In this manner, we were able to discard many combinations of input parameters. Table 2 gives the final range of possible input parameters which were able to simulate satisfactorily PO time-resolved traces and PO yields, indicating that BR lies between 20 to 50%. We therefore quote the BRs for reactions R1 and R2 as (35 ± 15)%. See the SI for further details of the Kintecus model (Ianni 2003) and of how the full range of input parameters given in Table 2 were reduced.

**Table 2.** Ranges of input parameters explored using Kintecus, and the possible ranges found after ruling out those that did not match the experimental PO traces.

| Input parameter | Range Explored | Possible Range |
|---|---|---|
| $^2$D:$^2$P Ratio | 0.22 – 18 | 1 – 3 |
| [$O_2$] / molecule cm$^{-3}$ | 0.5 – 3.0 × 10$^{14}$ | 1.0 – 1.5 × 10$^{14}$ |
| BR for P* + $O_2$ → PO + O | 25 – 100 % | 50 – 100 % |
| $^4$S:$^2$P Ratio | 0 – 18 | 0 – 2 |

## 4. Discussion

### *4.1 Stellar optical excitation of P($^4$S)*

We have shown here that the excited P* states are much more reactive with $H_2O$ and $H_2$ than ground-state P atoms. Similar behavior was observed previously for the reactions of P* and P with $O_2$ (Douglas et al. 2019). The potential importance of optical excitation of P($^4$S) in the stellar radiation field should therefore be considered. Although the optical transitions P($^4S_{3/2} \rightarrow {}^2D_{5/2}$) at 879 nm and P($^4S_{3/2} \rightarrow {}^2P_{1/2}$) at 534 nm are strongly forbidden with Einstein *A* coefficients of $2.0 \times 10^{-4}$ and $4.3 \times 10^{-2}$ s$^{-1}$, respectively (Kramida et al. 2021), the wavelengths at which these transitions occur are relatively close to the peak of the approximately black-body emission curves from AGB stars with effective surface temperatures below 4000 K (Gustafsson *et al.* 2008).

The transition oscillator strengths (Kramida et al. 2021) can be used to calculate the following temperature-dependent absorption cross sections (Okabe 1978): $\sigma(\mathrm{P}(^4S_{3/2} - {}^2D_{5/2})) = 6.6 \times 10^{-23} (T/1000)^{0.50}$ cm$^2$; $\sigma(\mathrm{P}(^4S_{3/2} - {}^2D_{3/2})) = 4.4 \times 10^{-23} (T/1000)^{0.50}$ cm$^2$; $\sigma(\mathrm{P}(^4S_{3/2} - {}^2P_{1/2}) = 1.1 \times 10^{-21} (T/1000)^{0.50}$ cm$^2$ and $\sigma(\mathrm{P}(^4S_{3/2} - {}^2P_{3/2}) = 5.3 \times 10^{-21} (T/1000)^{0.50}$ cm$^2$. The stellar irradiance flux from the MARCS data-base (Gustafsson et al. 2008) for an evolved star with $T_* = 3300$ K indicates a photon flux at 879 nm of $3.0 \times 10^{17}$ ph cm$^{-2}$ s$^{-1}$ nm$^{-1}$ at 2R$_*$. Hence the rate of optical excitation of P($^4$S → $^2$D) = $1.2 \times 10^{-7}$ s$^{-1}$, corresponding to an e-folding time $\tau$ = 98 days. The corresponding photon flux at 533 nm is $1.5 \times 10^{17}$ ph cm$^{-2}$ s$^{-1}$ nm$^{-1}$ at 2R$_*$, so the P($^4$S → $^2$P) rate of excitation = $2.08 \times 10^{-6}$ s$^{-1}$, with $\tau$ = 5.6 days. These rates are compared with collisional energy transfer rates in the next section.

### *4.2 A new chemical network for Phosphorus*

In this section we describe the development of the phosphorus chemistry network which is illustrated schematically in Figure 6. Table 3 lists the reactions in the network, along with the rate coefficients and a note of their source. In addition to the experimental results from the present study, we have previously measured a number of other rate coefficients for reactions involving P-bearing species (Douglas et al. 2019, Douglas et al. 2020). In order to estimate rate coefficients for reactions that have not been studied experimentally, we have used electronic structure theory at the CBS-QB3 level (Frisch *et al.* 2016, Montgomery et al. 2000) to determine the stationary points on the relevant reaction potential energy surfaces, combined with the Master Equation Solver for Multi-Energy well Reactions (MESMER) program (Glowacki *et al.* 2012). The use of these theoretical approaches is described in detail in the SI, along with notes on individual reactions. This approach should represent a significant improvement on our earlier study (Gobrecht et al. 2016), where rate coefficients for P-bearing species were set to their N analogues. In general, the new rate coefficients are within an order of magnitude of those assuming N↔P isovalence.

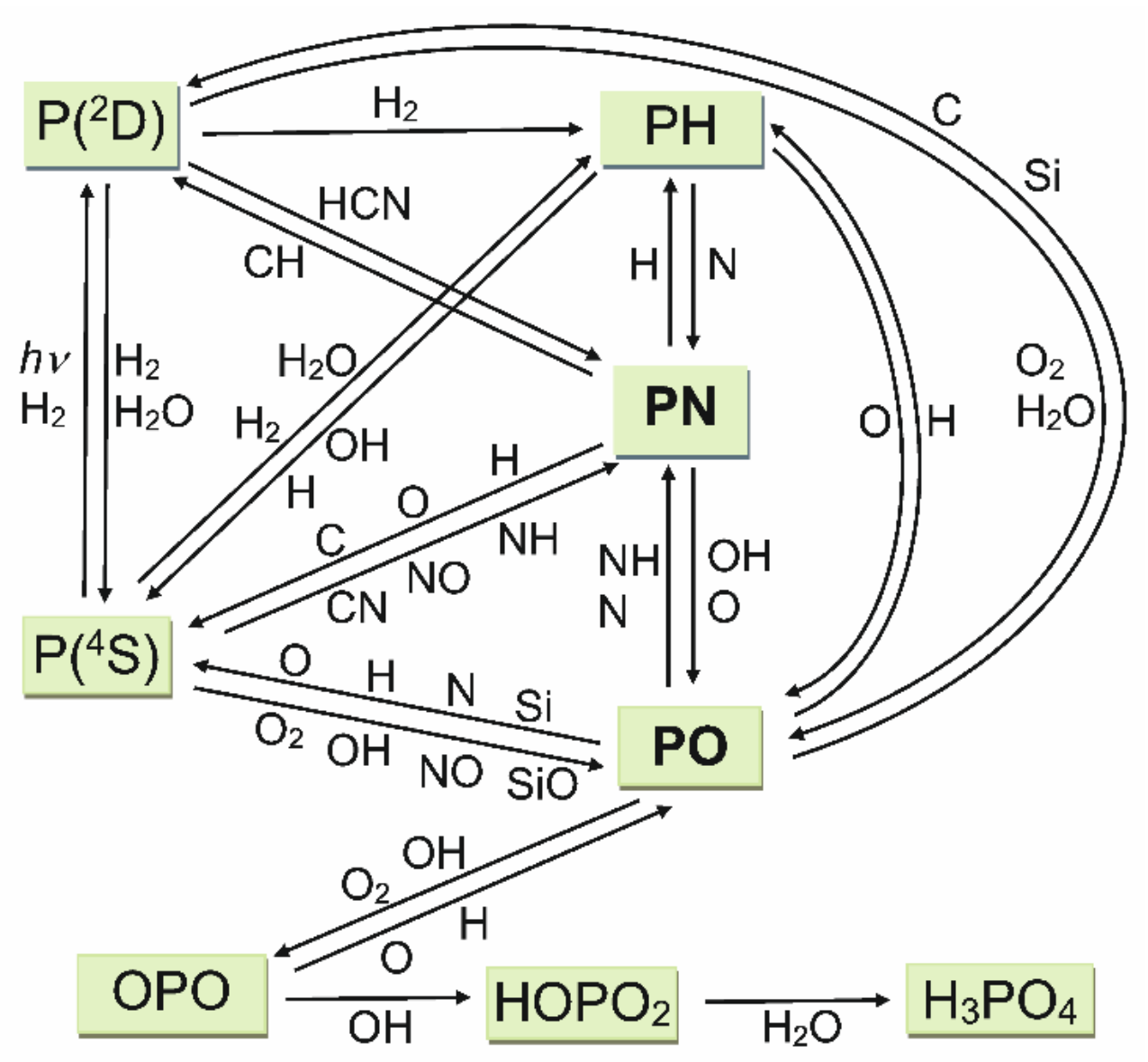

**Figure 6.** Schematic diagram of the phosphorus chemistry network developed in the present study.

The rate coefficient for collisional excitation of P($^4$S) to P($^2$D) ($k_{-21}$ in Table 3), shows that at a temperature of 1500 K and [$H_2$] = $10^{11}$ cm$^{-3}$ (i.e. typical values around 2$R_*$ in an outflow), the lifetime for collisional excitation is $1/(k_{-21}$ [$H_2$]) ~ 4300 s. This is 1950 times shorter than the optical pumping lifetime of 98 days (Section 4.1), and so optical excitation of P($^4$S → $^2$D) will not be significant in the inner stellar wind. In contrast, collisional excitation of P($^4$S) to P($^2$P) will occur with an e-folding lifetime of 37 days under the same conditions, which is significantly longer than the optical pumping lifetime of 5.6 days (Section 4.1). Nevertheless, since collisional excitation of P($^4$S) to P($^2$D) is 110 times faster than optical pumping to P($^2$P), the role of P($^2$P) should be very limited in this environment, and so reactions involving P($^2$P) are not included in the chemistry network (Table 3).

In the experiments on P($^2$D) + $H_2$, although we measured the temperature-dependent rate coefficient for the removal of P($^2$D), it was not possible to determine whether physical quenching to P($^4$S) or reaction to PH + H was occurring. The measured activation energy is 5.8 ± 0.2 kJ mol$^{-1}$. Calculations we performed at the very accurate W1BD level of theory (Barnes *et al.* 2009, Frisch et al. 2016) show that the reaction P($^2$D) + $H_2$ → PH + H is endothermic by 3.3 kJ mol$^{-1}$, which accords with the measured activation energy (within the expected error of 4 kJ mol$^{-1}$ at this level of theory). However, we also find that there is a barrier of 16 kJ mol$^{-1}$ on the doublet potential energy surface for the reaction. This barrier is significantly larger than the measured activation energy, which is good evidence that physical (i.e. collisional) quenching, rather than reaction, removes $P(^2D)$ in the presence of $H_2$. Nevertheless, as a sensitivity study in the stellar outflow modelling described in the next section, we treated unreactive and reactive collisions between P and $H_2$ as two extreme cases: the more likely *Scenario 1*, where $P(^2D) + H_2 \rightarrow P(^4S) + H_2$ and the reverse reaction are unreactive (reactions 21 and -21 in Table 3); or *Scenario 2*, where $P(^2D) + H_2 \rightarrow PH + H$, $P(^4S) + H_2 \rightarrow PH + H$ and the reverse $PH + H \rightarrow P(^4S) + H_2$ are reactive (reactions 21′, 22′ and -22′ in Table 3).

Lastly, we include in the network the formation of $HOPO_2$ and $H_3PO_4$ (phosphoric acid). The rate coefficients for the reactions which form these species (R19 and R20 in Table 3) are taken from a model that we developed recently to describe phosphorus chemistry in planetary atmospheres, where phosphate is a very stable sink for P produced by meteoric ablation (Plane *et al.* 2021).

**Table 3.** Reactions and rate coefficients in the Phosphorus chemical network

| | Reaction | Rate coefficient [a] | Notes |
|---|---|---|---|
| 1 | $P + O_2 \rightarrow PO + O$ | $4.2 \times 10^{-12} \exp(-600/T)$ | Douglas et al. (2019) |
| -1 | $PO + O \rightarrow P + O_2$ | $1.8 \times 10^{-13} (T/300)^{0.79} \exp(-10054/T)$ | Detailed balance with R1 |
| 2 | $PO + O_2 \rightarrow OPO + O$ | $2.3 \times 10^{-11} \exp(-100/T)$ | (Douglas et al. 2020) |
| -2 | $OPO + O \rightarrow PO + O_2$ | $3.5 \times 10^{-11} \exp(-1449/T)$ | Detailed balance with R2 |
| 3 | $P + NO \rightarrow PN + O$ | $7.1 \times 10^{-11} \exp(-6427/T)$ | MESMER calculation [b] |
| -3 | $PN + O \rightarrow P + NO$ | $1.0 \times 10^{-10} \exp(-3158/T)$ | MESMER calculation [b] |
| 4 | $P + OH \rightarrow PO + H$ | $6.9 \times 10^{-12} (T/300)^{-0.29}$ | Trajectory calculation [b] |
| -4 | $PO + H \rightarrow P + OH$ | $4.2 \times 10^{-13} (T/300)^{0.82} \exp(-16791/T)$ | Detailed balance with R4 |
| 5 | $P + H_2O \rightarrow PH + OH$ | $9.0 \times 10^{-10} \exp(-24319/T)$ | Detailed balance with R-5 |
| -5 | $PH + OH \rightarrow P + H_2O$ | $1.0 \times 10^{-10} (T/300)^{0.167}$ | Collision capture rate [b] |
| 6 | $PO + N \rightarrow P + NO$ | $5.5 \times 10^{-11} \exp(-12869/T)$ | MESMER calculation [b] |
| -6 | $P + NO \rightarrow PO + N$ | $1.0 \times 10^{-10} \exp(-17056/T)$ | MESMER calculation [b] |
| 7 | $PO + N \rightarrow PN + O$ | $9.8 \times 10^{-11} \exp(175/T)$ | MESMER calculation [b] |
| -7 | $PN + O \rightarrow PO + N$ | $2.7 \times 10^{-10} \exp(-744/T)$ | MESMER calculation [b] |
| 8 | $PH + O \rightarrow PO + H$ | $2.0 \times 10^{-10}$ | Collision capture rate [b] |
| -8 | $PO + H \rightarrow PH + O$ | $2.8 \times 10^{-10} (T/300)^{0.26} \exp(-34462/T)$ | Detailed balance with R8 |
| 9 | $PN + OH \rightarrow PO + NH$ | $2.6 \times 10^{-10} \exp(-13002/T)$ | Detailed balance with R-9 |
| -9 | $PO + NH \rightarrow PN + OH$ | $1.0 \times 10^{-10}$ | Collision capture rate [b] |
| 10 | $P + NH \rightarrow PN + H$ | $3.0 \times 10^{-10}$ | Collision capture rate [b] |
| -10 | $PN + H \rightarrow P + NH$ | $4.3 \times 10^{-12} (T/300)^{1.19} \exp(-27868/T)$ | MESMER calculation [b] |
| 11 | $PN + CH \rightarrow HCN + P(^2D)$ | $1.0 \times 10^{-10}$ | Collision capture rate [b] |
| -11 | $HCN + P(^2D) \rightarrow PN + CH$ | $1.0 \times 10^{-10}$ exp(-39200) | Detailed balance with R11 |
| 12 | $PN + C \rightarrow CN + P$ | $1.1 \times 10^{-10}$ | Collision capture rate [b] |
| -12 | $P + CN \rightarrow PN + C$ | $7.3 \times 10^{-10}$ exp(-16240) | Detailed balance with R12 |
| 13 | $PO + C \rightarrow P(^2D) + CO$ | $5.0 \times 10^{-11}$ | See the SI [b] |
| 14a | $PO + Si \rightarrow P + SiO$ | $2.1 \times 10^{-10} (T/300)^{0.10}$ | MESMER calculation [b] |
| 14b | $PO + Si \rightarrow P(^2D) + SiO$ | $2.1 \times 10^{-10} (T/300)^{0.01}$ | MESMER calculation [b] |
| -14a | $P + SiO \rightarrow PO + Si$ | $3.5 \times 10^{-10}$ exp(-24416/$T$) | MESMER calculation [b] |
| 15 | $PH + N \rightarrow PN + H$ | $1.0 \times 10^{-10} (T/300)^{0.17}$ | MESMER calculation [b] |
| -15 | $PN + H \rightarrow PH + N$ | $3.1 \times 10^{-10} (T/300)^{0.51}$ exp(-35427/$T$) | MESMER calculation [b] |
| 16 | $PO + OH \rightarrow OPO + H$ | $1.2 \times 10^{-10}$ | Collision capture rate [b] |
| -16 | $OPO + H \rightarrow PO + OH$ | $1.0 \times 10^{-9}$ exp(-9490/$T$) | Detailed balance with R-16 |
| 17a | $P(^2D) + H_2O \rightarrow PO + H_2$ | $1.6 \times 10^{-11}$ | This study (reaction 2a [c]) |
| 17b | $P(^2D) + H_2O \rightarrow P + H_2O$ | $3.0 \times 10^{-11}$ | This study (reaction 2b [c]) |
| 18 | $P(^2D) + O_2 \rightarrow PO + O$ | $1.2 \times 10^{-11} (T/300)^{0.82}$ exp(177/$T$) | Douglas et al. (2019) |
| 19 | $OPO + OH (+ H_2) \rightarrow HOPO_2$ | $2.2 \times 10^{-26} (T/300)^{-5.25}$ | Plane et al. (2021) |
| 20 | $HOPO_2 + H_2O (+ H_2) \rightarrow H_3PO_4$ | $3.0 \times 10^{-26} (T/300)^{-7.53}$ | Plane et al. (2021) |
| **Scenario 1** | | | |
| 21 | $P(^2D) + H_2 \rightarrow P + H_2$ | $8.0 \times 10^{-11}$ exp(-699/$T$) | This study (reaction R4b [c]) |
| -21 | $P + H_2 \rightarrow P(^2D) + H_2$ | $2.0 \times 10^{-10}$ exp(-17056/$T$) | Detailed balance with R21 |
| **Scenario 2** | | | |
| 21′ | $P(^2D) + H_2 \rightarrow PH + H$ | $8.0 \times 10^{-11}$ exp(-699/$T$) | This study (reaction R4a [c]) |
| 22′ | $P + H_2 \rightarrow PH + H$ | $6.7 \times 10^{-10}$ exp(-17349/$T$) | Transition state theory [b] |
| -22′ | $PH + H \rightarrow P + H_2$ | $6.7 \times 10^{-11}$ exp(-346/$T$) | Detailed balance with R22′ |

[a] Units: $cm^3$ $molecule^{-1}$ $s^{-1}$ for bimolecular reactions; $cm^6$ $molecule^{-2}$ $s^{-1}$ for termolecular reactions.
[b] See the Supporting Information for further details.
[c] Reaction number in the text.

### *4.3 Numerical modelling*

We use here a model that we applied recently to circumstellar gas trajectories, including non-pulsating and pulsating outflows, for two late-type stars with a different evolutionary stage: a semi-regular (SRV) AGB star, and a Mira-type (MIRA) AGB star (Gobrecht *et al.* 2022). In the present study, we use the gas trajectories of the MIRA model star in oxygen-rich conditions (C/O = 0.75), characterised by a total density of $n_{gas} = 4 \times 10^{14}$ $cm^{-3}$ and a temperature of 2000 K at the stellar surface. The phosphorus chemistry in Table 3 was added to the oxygen-rich kinetic network in the model. The resulting model outputs are illustrated in Figure 7. Note that because the recombination reaction R20 of $HOPO_2$ with $H_2O$ is much faster than the recombination R19 of OPO with OH (since [$H_2O$] >> [OH]), $HOPO_2$ is rapidly converted to $H_3PO_4$ and, for reasons of clarity, this latter species is not shown in the panels in Figure 7. The total gas density, temperature and fractional abundances of species in the outflow which impact on the phosphorus chemistry (H, $H_2$, $H_2O$, OH, O, $O_2$, N and $N_2$) are listed in Table S5 in the SI.

In the non-pulsating case (Figure 7a), the time for the outflow to reach 1.5 $R_*$ is 377 days, and 551 days to reach 2 $R_*$. The corresponding conditions in the gas are $n_{gas} = 5.8 \times 10^{12}$ cm$^{-3}$ and $T = 1570$ K at 1.5 $R_*$ and $4.31 \times 10^{11}$ cm$^{-3}$ and 1320 K at 2.0 $R_*$. In this case, the phosphorus chemistry is dominated by PO, which is mainly formed by the barrierless reactions P + OH (R4) and P($^2$D) + $H_2O$ (R17a). This is because in the hot region OH has a relatively high concentration, and the P($^2$D) / P($^4$S) ratio is favoured by collisional excitation of P($^4$S) by $H_2$ (reaction R21). The reaction of P with $O_2$ (R1) is of secondary importance because of the low $O_2$ density in the inner wind of the outflow.

Although the predicted PO fractional abundance greater than $10^{-7}$ is a dramatic improvement on our previous model (where PO was under-predicted by around 3 orders of magnitude (Gobrecht et al. 2016)), in this non-pulsating model the PN abundance is just over $10^{-9}$, which is roughly 2 orders of magnitude lower than observed (see Section 1). Moreover, only small amounts of OPO and $HOPO_2$ are simulated, so the formation of the potential biomolecule $H_3PO_4$ is limited. Scenario 2 makes very little difference to PO and PN, though there is a significant increase in PH and a decrease in P, as expected when collisions between P($^2$D and $^4$S) with $H_2$ are treated as being reactive (Table 3).

In the pulsating model (Figure 7b), the hydro-dynamical timescale is set by the pulsation period of 450 days and corresponds to the time for the non-pulsating outflow to reach a radial position between 1.5 and 2 $R_*$. We assume that the stellar pulsation steepens into a shock wave at the stellar surface. The immediate post-shock conditions are very hot and dense ($n_{gas} = 6.8 \times 10^{15}$ cm$^{-3}$ and $T = 4480$ K). In this case, PN represents the most abundant phosphorus-bearing molecule for phases $\Phi > 0.5$ in the model. However, the abundance of PO does not persist in the pulsating model at 1 $R_*$, and is comparable to observations only when $\Phi$ lies between 0.25 and 0.55.

The difference in the phosphorus chemistry between the non-pulsating and pulsating models arises because in the hotter and denser pulsating model, a much larger fraction of molecular $N_2$ becomes thermally dissociated:

$$N_2 + M \rightarrow N + N + M \quad (M = H_2, H, He\ etc.) \qquad R23$$

where R23 has a substantial activation energy: $k_{23} = 9.2 \times 10^{-5}\ (T/298)^{-2.50} \exp(-113060/T)$ cm$^3$ molecule$^{-1}$ s$^{-1}$ (Kewley & Hornung 1974). PO can then react with atomic N to form PN via reaction R7 (Table 3). This reaction is close to thermoneutral ($\Delta H^\circ_{(0\ K)} = -7$ kJ mol$^{-1}$), so the PO/PN ratio is largely controlled by the O/N ratio. In the cooler non-pulsating model, R23 is very slow and the O/N ratio is therefore larger.

Interestingly, a significant quantity of $HOPO_2$ (and $H_3PO_4$) forms, with the fractional abundance of $4 \times 10^{-9}$ exceeding the corresponding abundance of the non-pulsating model by factor of ~20. This arises because of a 20-fold increase in $O_2$ after the shock (Table S5), which favours conversion of PO into OPO (R2); significant reductions early in the pulsation cycle ($\Phi > 0.2$) in O and H, which destroy OPO (R-2 and R-16); and the 15-fold increase in total gas density which favours the termolecular recombination of OPO with OH to form $HOPO_2$.

As a final modelling step, we combine the non-pulsating and the pulsating model by assuming a monotonic outflow from the stellar surface (1 $R_*$) out to 1.5 $R_*$ ($n_{gas} = 5.8 \times 10^{13}$ $cm^{-3}$ and $T = 1570$ K), followed by a pulsation-induced shock. This results in immediate post-shock conditions of $n_{gas} = 3.8 \times 10^{13}$ $cm^{-3}$ and $T = 3360$ K at 1.5 $R_*$, which then relaxes over the timescale of the pulsation period. In the post-shock gas model the PO and PN abundances are around $10^{-7}$ for the intermediate phases (see Figure 7c). At full phase ($\Phi \rightarrow 1$) the PO abundance ($\sim 2 \times 10^{-9}$) is rather low, but still orders of magnitude higher than in the pulsating model at 1 $R_*$. The model therefore confirms earlier work (e.g. Lefloch et al. (2016) and Mininni et al. (2018)) which proposed that high PN levels form in the shocked regions of astrophysical outflows. This probably at least partially explains why the PO/PN ratios observed around different stars are quite variable (see Section 1). A final point of note is that in this model run $H_3PO_4$ (or $HOPO_2$) forms effectively in the post-shock gas and reaches abundances above $10^{-8}$.

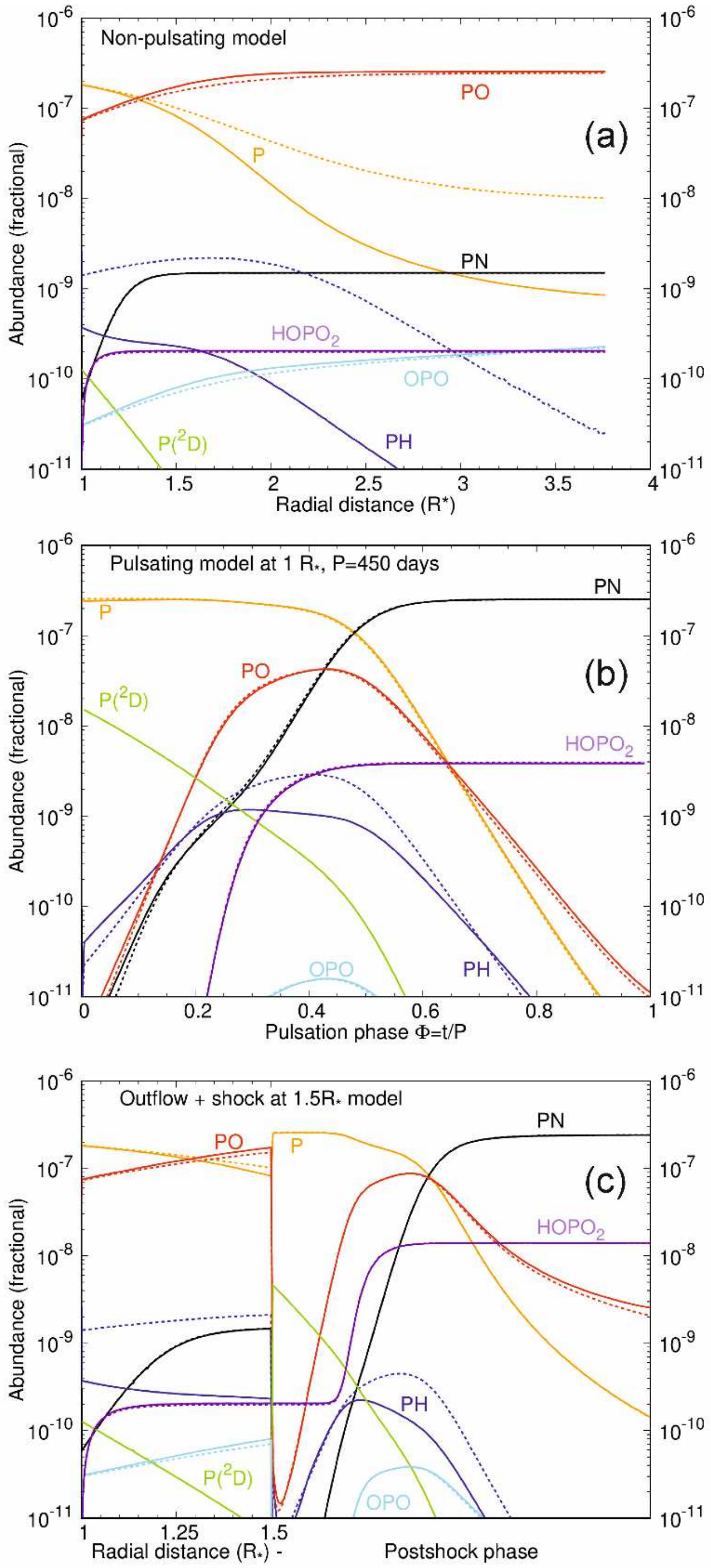

**Figure 7.** Molecular abundances of the P-bearing species in the different stellar outflow models: (a) a non-pulsating, steady outflow with a monotonically increasing radial distance from the star, corresponding to decreasing gas densities and temperatures; (b) the post-shock gas of a pulsating outflow that experiences a shock at 1 R* and relaxes to its preshock condition within one pulsation period; (c) a combination of a steady outflow (from 1.0-1.5 $R_*$) and a subsequent pulsational shock at 1.5 $R_*$ including the related post-shock gas. Solid lines correspond to Scenario 1 and dashed lines to Scenario 2 (see text). Note that $HOPO_2$ is rapidly converted to $H_3PO_4$, and so the latter is not shown for clarity.

## 5. Conclusions

In this study of phosphorus chemistry in stellar outflows we first investigated the role of excited P atom chemistry in producing PO from the reaction of the low-lying metastable P($^2$D) and P($^2$P) states with $H_2O$. The measured rate coefficients of both reactions are reasonably fast, with a PO yield of ~35%. $H_2$ was shown to react with both P states relatively efficiently, with physical quenching, rather than chemical reaction to produce PH + H, being the more likely pathway for P($^2$D) + $H_2$. We then developed a comprehensive phosphorus chemistry network for inclusion in a stellar outflow model, using a combination of electronic structure theory calculations and a Master Equation treatment of reactions taking place over complex potential energy surfaces. The new model shows that at high temperatures within ~2 stellar radii, collisional excitation of ground-state P($^4$S) to P($^2$D), followed by reaction with $H_2O$, is a significant pathway for producing PO, along with the reaction between P($^4$S) and OH. The model also demonstrates that the PN fractional abundance in a steady (non-pulsating) outflow is under-predicted by about 2 orders of magnitude compared with observation. However, under shocked conditions where enough thermal dissociation of $N_2$ occurs at temperatures above 4000 K, the resulting N atoms convert a substantial fraction of PO to PN, which is consistent with reports that PN tends to correlate in some (often carbon-rich) environments with the shock tracer SiO.

## Acknowledgements

This study was supported by funding from the UK Science and Technology Facilities Council (grant ST/P000517/1). D.G. was funded by the project grant "The Origin and Fate of Dust in Our Universe" from the Knut and Alice Wallenberg Foundation.

## Data availability

All the experimental and modelling data is included in Tables in the main paper or the Supporting Information.

## References

Acuña A.U., Husain D., Wiesenfeld J.R., 1973, J. Chem. Phys., 58, 494

Asplund M., Grevesse N., Sauval A.J., Scott P., 2009, in: Blandford R., Kormendy J., van Dishoeck E. eds., Ann. Rev. Astron. Astrophys. Annual Reviews, Palo Alto, p. 481

Barnes E.C., Petersson G.A., Jr. J.A.M., Frisch M.J., Martin J.M.L., 2009, J. Chem. Theor. Comput., 5, 2687

Bernal J.J., Koelemay L.A., Ziurys L.M., 2021, Astrophys. J., 906, art. no.: 55

Bridgeman O.C., Aldrich E.W., 1964, J. Heat Transfer, 86, 279

Chantzos J., Rivilla V.M., Vasyunin A., Redaelli E., Bizzocchi L., Fontani F., Caselli P., 2020, Astron. Astrophys., 633, art. no.: A54

De Beck E., Kamiński T., Patel N.A., Young K.H., Gottlieb C.A., Menten K.M., Decin L., 2013, Astron. Astrophys., 558, art. no.: A132

Douglas K.M., Blitz M.A., Mangan T.P., Plane J.M.C., 2019, J. Phys. Chem. A, 123, 9469

Douglas K.M., Blitz M.A., Mangan T.P., Western C.M., Plane J.M.C., 2020, J. Phys. Chem. A, 124, 7911

Fontani F., Rivilla V.M., van der Tak F.F.S., Mininni C., Beltran M.T., Caselli P., 2019, Mon. Not. Roy. Astron. Soc., 489, 4530

Frisch M.J., et al., 2016. Gaussian, Inc., Wallingford, CT, USA,

Glowacki D.R., Liang C.-H., Morley C., Pilling M.J., Robertson S.H., 2012, J. Phys. Chem. A, 116, 9545−9560

Gobrecht D., Cherchneff I., Sarangi A., Plane J.M.C., Bromley S.T., 2016, Astron. Astrophys., 585, art. no.: A6

Gobrecht D., Plane J.M.C., Bromley S.T., Decin L., Cristallo S., Sekaran S., 2022, Astron. Astrophys., 658, art. no. A167

Gómez Martín J.C., Blitz M.A., Plane J.M.C., 2009, Phys. Chem. Chem. Phys., 11, 671

Gustafsson B., Edvardsson B., Eriksson K., Jørgensen U.G., Nordlund Å., Plez B., 2008, Astron. Astrophys., 486, 951

Ianni J.C., 2003, in: Bathe K.J. ed., Computational Fluid and Solid Mechanics. Elsevier Science Ltd., Oxford,

Jimenez-Serra I., Viti S., Quenard D., Holdship J., 2018, Astrophys. J., 862, art. no.: 128

Kewley D.J., Hornung H.G., 1974, Chem. Phys. Lett., 25, 531

Kramida A., Ralchenko Y., Reader J., 2021, NIST Atomic Spectra Database (version 5.9) [Online]. National Institute of Standards and Technology, Gaithersburg, MD.

Lefloch B., et al., 2016, Mon. Not. Roy. Astron. Soc., 462, 3937

Mangan T.P., McAdam N., Daly S.M., Plane J.M.C., 2019, J. Phys. Chem. A, 123, 601

Mininni C., Fontani F., Rivilla V.M., Beltran M.T., Caselli P., Vasyunin A., 2018, Mon. Not. Roy. Astron. Soc., 476, L39

Molpeceres G., Kästner J., 2021, Astrophys. J., 910, art. no.: 55

Montgomery J.A., Frisch M.J., Ochterski J.W., Petersson G.A., 2000, J. Chem. Phys., 112, 6532

Okabe H., 1978, Photochemistry of Small Molecules. John Wiley & Sons, New York

Plane J.M.C., Feng W.H., Douglas K.M., 2021, J. Geophys. Res.-Space Phys., 126, art. no.: e2021JA029881

Rivilla V.M., et al., 2018, Mon. Not. Roy. Astron. Soc., 475, L30

Sausa R.C., Miziolek A.W., Long S.R., 1986, J. Phys. Chem., 90, 3994

Schwartz A.W., 2006, Phil. Trans. R. Soc. B, 361, 1743

Sil M., et al., 2021, Astronom. J., 162, art. no.: 119

Souza A.C., Silva M.X., Galvao B.R.L., 2021, Mon. Not. Roy. Astron. Soc., 507, 1899

Velilla Prieto L., et al., 2017, Astron. Astrophys., 597, art. no.: A25

Walton C.R., et al., 2021, Earth-Science Reviews, 221

Williams D.A., Hartquist T.W., 2013, The Cosmic-Chemical Bond: Chemistry from the Big Bang to Planet Formation. Royal Society of Chemistry, Cambridge, U.K.

Ziurys L.M., Schmidt D.R., Bernal J.J., 2018, Astrophys. J., 856, art. no. 169

**Figure Captions**

**Figure 1.** PO($A^2\Sigma^+$) fluorescence signal captured from the oscilloscope from experiments using $N_2$ as a bath gas, at three [$H_2O$]: black squares [$H_2O$] = $1.8 \times 10^{16}$ molecule cm$^{-3}$; red circles [$H_2O$] = $1.1 \times 10^{16}$ molecule cm$^{-3}$; blue triangles [$H_2O$] = $0 \times 10^{16}$ molecule cm$^{-3}$. The solid lines are single exponential fits to each signal decay; these fits start far enough in time after the respective peaks in the bi-exponential signal that the early fast growth has ended. Inset: PO($A^2\Sigma^+$) fluorescence lifetime, $k_f$, *vs* [$H_2O$], from experiments using $N_2$ as a bath gas.

**Figure 2.** P($^2$D) LIF signal following PLP of $PCl_3$ at a total pressure of 7.64 Torr and [$H_2O$] = $7.19 \times 10^{14}$ molecule cm$^{-3}$, at $T$ = 593 K. Inset: a bimolecular plot for reaction R2 at $T$ = 593 K, giving $k_2 = (4.99 \pm 0.30) \times 10^{-11}$ cm$^3$ molecule$^{-1}$ s$^{-1}$.

**Figure 3.** Temperature dependence of the rate coefficient for a) P($^2$P) + $H_2$, b) P($^2$D) + $H_2$, c) P($^2$P) + $H_2O$, and d) P($^2$D) + $H_2O$. Black open squares are from this study, with Arrhenius fits indicated by red lines; solid blue circles are from Acuña et al. (1973).

**Figure 4.** Top panels: PO LIF signals following PLP of $PCl_3$ at various [$H_2O$]. Solid lines are fits of Equation 5 to the data. Middle panels: Simulated and experimental PO profiles at high $H_2O$ ([$H_2O$] = $6 \times 10^{15}$ (left side) and [$H_2O$] = $1 \times 10^{15}$ (right side)). Bottom panels: Simulated and experimental PO profiles at zero $H_2O$. All left-hand panels are for experiments in He, while all right-hand panels are for experiments in $N_2$. Simulated fits which are considered good are (poor fits and reasons in brackets): c) red upward triangles (blue squares rise too fast, green diamonds loss too slow); d) red upward triangles (blue squares loss too slow, green diamonds rise too fast and loss too slow); e) red upward triangles (blue squares rise too fast); f) red upward triangles and yellow downward triangles (blue squares loss too fast, green diamonds rise and loss too slow).

**Figure 5.** Experimental (open dark red stars) and simulated (closed symbols) PO yields. Top panel: experiments carried out in He, with model input parameters of $^2$D/$^2$P = 2:1, [$O_2$] = $1 \times 10^{14}$ cm$^{-3}$, $O_2$ BR = 100%, $^4$S:$^2$P = 0:1. Bottom panel: experiments carried out in $N_2$, with model input parameters of $^2$D/$^2$P = 2:1, [$O_2$] = $1 \times 10^{14}$ cm$^{-3}$, $O_2$ BR = 50%, $^4$S:$^2$P = 2:1.

**Figure 6.** Schematic diagram of the phosphorus chemistry network developed in the present study

**Figure 7.** Molecular abundances of the P-bearing species in the different stellar outflow models: (a) a non-pulsating, steady outflow with a monotonically increasing radial distance from the star, corresponding to decreasing gas densities and temperatures; (b) the post-shock gas of a pulsating outflow that experiences a shock at 1 R* and relaxes to its preshock condition within one pulsation period; (c) a combination of a steady outflow (from 1.0-1.5 $R_*$) and a subsequent pulsational shock at 1.5 $R_*$ including the related post-shock gas. Solid lines correspond to Scenario 1 and dashed lines to Scenario 2 (see text). Note that $HOPO_2$ is rapidly converted to $H_3PO_4$, and so the latter is not shown for clarity.